\documentstyle[preprint,aps,eqsecnum,epsfig]{revtex} 
\newcommand{\be}{\begin{equation}}
\newcommand{\ee}{\end{equation}}
\newcommand{\bea}{\begin{eqnarray}}
\newcommand{\eea}{\end{eqnarray}}
\newcommand{\dbark}{\frac{d^3k}{(2\pi)^3}}

\tighten 
\begin{document} 
\title{\bf THE SPECIFIC HEAT OF NORMAL, DEGENERATE QUARK MATTER: NON-FERMI LIQUID CORRECTIONS}   
\author{\bf D. Boyanovsky$^{(a,b)}$, H. J. de Vega$^{(b,a)}$}
\address{(a) Department of Physics and Astronomy, University of 
Pittsburgh, Pittsburgh  PA. 15260, U.S.A\\(b) LPTHE, Universit\'e Pierre et Marie Curie (Paris VI) et Denis Diderot 
(Paris VII), Tour 16, 1er. \'etage, 4, Place Jussieu, 75252 Paris, Cedex 05,France}
\date{\today}
\maketitle 
\begin{abstract} 
In normal degenerate quark matter, the exchange of dynamically screened transverse gluons introduces infrared divergences
in the quark self-energies that lead to the breakdown of the Fermi liquid description. If the core of neutron stars
are composed of quark matter with a normal component, cooling by direct quark Urca processes may be modified by non-Fermi
liquid corrections.
 We find that while the quasiparticle density of states is finite and non-zero at the Fermi surface, its frequency derivative diverges and results in non-Fermi liquid corrections to the specific heat of the normal, degenerate component of quark matter.  We study these non-perturbative non-Fermi liquid corrections to the specific heat and the temperature dependence of the chemical potential  and
show that these lead to a reduction of the specific heat.

\end{abstract} 

\section{Introduction and Motivation} 

A detailed description of the phase diagram of QCD in hot and dense environments can lead to a deeper understanding
of Early Universe cosmology as well as the physics of compact stellar objects. Current experimental programs at CERN-SPS and BNL-RHIC probe QCD at high temperatures but relatively small baryon (and quark) density\cite{muzi} and the future ALICE program at
CERN-LHC will probe at even higher temperatures and smaller densities, in conditions that prevailed a few microseconds after
the Big Bang. The goal of these ultrarelativistic heavy ion experiments is to search for the quark gluon plasma, a novel
phase of QCD in which quarks and gluons are deconfined. The region of the phase diagram of QCD for low temperatures
$T \leq 10 ~\mbox{Mev}$ but densities a few times larger than that of nuclear matter, cannot be studied with terrestrial
accelerators but are the realm of compact stellar objects: neutron stars and pulsars. There is the tantalizing possibility that at the core of neutron stars where the density is up to 5 times that of nuclear matter, there could be a component of
cold, degenerate quark matter with temperatures $T \leq 1~\mbox{Mev}$ and chemical potentials $\mu \sim 300-500 ~\mbox{Mev}$\cite{glendenning,weber}. 

An important aspect of dense QCD is the possibility of color superconductivity\cite{bailin}-\cite{reviews}. One gluon
exchange leads to a pairing instability of the quark Fermi surface in the color  antitriplet channel\cite{bailin}, and
while originally superconducting gaps were estimated to be of order $\Delta \sim 1~\mbox{Mev}$ more recent estimates
including dynamical screening of the exchanged gluon\cite{son} conclude that the gaps could be $\Delta \sim 50-100 ~\mbox{Mev}$\cite{alford}-\cite{reviews}. 

The existence of quark matter phases in the cores of neutron stars and pulsars could have distinct observational
consequences. The equation of state of quark matter may lead to pronounced delays in the spin-up history of neutron stars
in  low mass
X-ray binaries and may explain recently observed anomalous frequency distributions\cite{glenweber}, a deconfinement transition can lead to observational consequences in the rotational properties of pulsars\cite{blaschke}, superconducting quark matter can influence the strength and distribution of magnetic
fields of pulsars\cite{blas,alfmag} and could influence the cooling history of young neutron stars\cite{quarksup} and protoneutron stars\cite{reddy,markrajarevius}. 

The properties of neutron stars and pulsars are studied by a satellite program that includes the Einstein Observatory,
ROSAT, AXAF, RXTE and more recently CHANDRA and XMM that measure the (soft) X-ray emission from neutron stars, thus studying their
cooling history as well as  rotational properties (spin-up and spin-down). 

A few seconds after the collapse of a Type II supernovae the newly born neutron star cools via the emission of neutrinos
and antineutrinos and within few minutes the temperature falls to about $10^9 K$ ($\sim 0.1~\mbox{Mev}$) cooling proceeds
via neutrino emission for another $10^{5-6}$ years before photon emission becomes the most important cooling
 mechanism\cite{shapiro}-\cite{page}.

The most efficient neutrino emission mechanisms are the direct URCA processes $f_1+ \l \rightarrow f_2+\nu_l~;~ f_2\rightarrow f_1+l+\bar{\nu}_l$ with $f_i$ being either baryons or quarks and $l$ either electrons or muons with
$\nu_l~;~\bar{\nu}_l$ their respective neutrinos and antineutrinos\cite{quarksup}. In the region between the crust and
the core, where nuclear matter is the dominant component this process occurs with  baryons and corresponds to the beta decay of the neutron and electron capture in a
proton. If  the core is composed of  quark matter the corresponding processes are direct quark Urca, $d\rightarrow u+e^-+\bar{\nu}_e ~ ; ~ u+e^- \rightarrow d+\nu_e$ which have been studied in detail in ref.\cite{iwamoto}. The cooling rate is determined by the equation\cite{shapiro}-\cite{page}

\be
\frac{d U}{dt} = C_v \frac{dT}{dt} = -(L_{\nu}+L_{\gamma})\label{cooling}
\ee

\noindent with $C_v$ the specific heat (at constant volume and baryon number) and $L_{\nu,\gamma}$ are the neutrino and
photon luminosities,  any other potential sources of energy loss such as magnetic field decay, differential 
rotation etc. had been neglected in the above equation.

The superconducting (and superfluid) component gives a contribution to the specific heat and the neutrino emissivity that is exponentially
 suppressed by the gap $\propto e^{-\Delta/T}$  if $\Delta \gg T$\cite{pethick,yakovlev,quarksup}. Recently, however, it was noted that novel aspects of cooling during the {\em protoneutron} star stage, when $T \sim
30-50 ~\mbox{Mev}$ could be a result of the color superconducting transition\cite{reddy} since in this regime 
$\Delta \leq T$. 

For two light 
flavors (u,d) of quarks one gluon exchange leads to an instability in the antitriplet channel leading to a color superconducting state known as (2SC)\cite{reviews,markrajarevius}. This state leaves quarks of one color ungapped while the others pair up with a gap $\Delta \sim 50-100~\mbox{Mev}$. The temperature of young neutron stars is $ T \sim 0.1-1 \mbox{Mev}$, hence in a color superconducting state
the paired quarks will not contribute to cooling. Including the strange quark the color superconducting state for three
flavors is a color-flavor-locked phase with all quarks paired.

 If the strange quark mass is too large to allow for S-wave 
pairing in the color-flavor-locked phase it may still form a  P-wave condensate but with a very small gap\cite{schaf}, 
in which case S-wave color superconductivity will be in the form of a (2SC) condensate. In this case the only color degree
of freedom remaining ``normal''  and the strange quark contribute to the cooling via neutrino emissivity but the
contribution from the strange quark will be suppressed by the (small) gap\cite{schaf} as well as by the small Cabibbo
angle\cite{iwamoto}. 


There are two important quantities that determine the cooling rate: the specific heat $C_v$ and the neutrino emissivity. Both are completely determined by the physics near  the Fermi surface of the degenerate quarks and leptons (the only important ones are electrons). 

 Iwamoto\cite{iwamoto} calculated the neutrino emissivity of normal quarks via the direct quark Urca processes, assuming
the validity of Fermi liquid theory, and found it  
to be $\epsilon \propto T_9^6$ ($T_9 = T/10^9 K$). The power of temperature is found on physical grounds: each degenerate
fermion (quarks and electron) has associated a phase space factor of  $ d^3 p/|p| \sim p_F T$ since only states within a region of
width $T$ near the Fermi surface
contribute, this gives three powers of T, energy-momentum conservation restricts the neutrino to have energy-momentum $\sim T$ and since the neutrinos are not Pauli blocked  their available phase space is $\sim E^2_{\nu} dE_{\nu} \propto T^3$. The energy conservation delta function leads to a factor $1/T$ which is compensated by a factor $E_{\nu}\sim T$ in the numerator for the energy loss.  Thus the energy loss by neutrino emission is $\propto \Pi_{f=u,d,e} (p_F(f)T) T^3$\cite{iwamoto}. The linear power of $T$ arising from phase space for each degenerate fermion is determined
by the single particle density of states near the Fermi surface which also determines the linear temperature dependence of the specific heat, as can be seen from the following argument.
The contribution from massless degenerate quarks to the temperature change in the internal energy  is determined by the states in a region of width $\approx T$ from the Fermi surface, within
which quarks have thermal energy $\approx T$. The density of single particle states at the Fermi surface $p_F=\mu$ for 
relativistic fermions is $\propto p^2_F = \mu^2$, thus within a region of width $T$ around the Fermi surface there are $\sim \mu^2 T$ states of energy $\sim T$  leading to their contribution to the temperature change in internal energy

\begin{equation}
\delta U_q(\mu,T) \propto \mu^2 T^2 \rightarrow C_v \propto \mu^2 T
\end{equation}

The main reason for repeating these arguments is to highlight that both the neutrino emissivity and the specific heat
of the normal component are determined by the (quasi) particle states within a region of width $T$ near the Fermi surface. 

In ref.\cite{son,ren} it was realized that the exchange of dynamically screened gluons that lead to the pairing instability also leads to a breakdown of the Fermi liquid picture for quarks near the Fermi surface. In particular these non-Fermi liquid corrections are responsible for 
a decrease in the gap\cite{ren}. A detailed study of non-Fermi liquid aspects of the normal state was recently presented in\cite{nflqcd}. There the  spectral density, dispersion relation and width of quasiparticles with momenta near the Fermi surface were derived at $T=0$ by 
implementing a renormalization group resummation of the leading logarithmic infrared divergences associated with the
emission of soft  dynamically screened transverse gluons. These non-Fermi liquid corrections modify the properties of quasiparticle
states near the Fermi surface.  Since cooling of degenerate quark matter is sensitive to the properties of the (quasi) particle states near the Fermi surface, it is natural to study the possible  non-Fermi liquid corrections to the specific heat as well as the neutrino emissivity.

Our goal in this article is to study in detail the non-Fermi liquid corrections to the specific heat to establish if these
result in an enhancement or suppression of the cooling rate via neutrino emission through direct quark Urca processes, postponing  the study of non-Fermi liquid effects upon the neutrino emissivity to a forthcoming article.

\section{Quasiparticle density of states, chemical potential and specific heat:}

Our strategy to obtain the specific heat (at constant volume and baryon density) is to first obtain the internal energy and take its temperature
derivative. The internal energy in turn is the expectation value of the full Hamiltonian $H=H_q+H_g+H_{q-g}$,  the sum of the
quark, gluon and interaction parts. The regime of interest for neutron star phenomena is 

\begin{equation}
T \approx 0.1-1~ \mbox{Mev} \; \; ; \; \; \mu_q \approx 0.3-0.5~ \mbox{Gev} 
\end{equation}

\noindent i.e, $T\ll \mu$. Furthermore the scale that enters in the resummed quark propagator\cite{nflqcd} is $M \approx g\mu/2\pi$ at energy scales $\sim 1~ \mbox{Gev}$ the strong coupling constant $\alpha_s \approx 0.6$ hence $M \sim \mu_q $ implying the hierarchy $T \ll M \sim \mu$. The contribution from massless gluons or Goldstone bosons remaining from the
superconducting transition, to the internal energy is 

\begin{equation}
U_g(T) \propto T^4
\end{equation}

Therefore the gluon contribution to the internal energy is subleading by a factor $T^2/\mu^2$ as compared to the quark contribution which, as argued above is $\delta U\sim \mu^2 T^2$. Hence
we will neglect the contribution from $H_g$ to the internal energy. The expectation value of $\tilde{H}=H_q+H_{q-g}$ can be obtained straightforwardly
from the spectral density of the quark field by the following observation: the quark Hamiltonian including the interaction with the gluon
field is bilineal in the quark field, and the equation of motion is
\begin{equation}
i \frac{\partial \Psi}{\partial t} = \frac{\delta \tilde{H}}{\delta {\Psi^{\dagger}}} 
\end{equation}

\noindent therefore it follows that

\begin{equation}
\int d^3x \bar{\Psi}(\vec x,t)~ i\gamma^0~\partial_t \Psi(\vec x,t) = \tilde{H}
\end{equation}

Hence we write

\begin{equation}
U(T,\mu) = \langle \tilde{H} \rangle = \int d^3 x~ i\partial_t \left[\langle\bar{\Psi}_b(\vec x',t')\Psi_a(\vec x,t)  \rangle \gamma^0_{b,a}  \right]_{\vec x'=\vec x;t=t'} 
\end{equation}

The fermion correlator can be written in terms of the spectral representation of its spatial Fourier transform
as follows\cite{nflqcd}

\bea
\langle\bar{\Psi}_b(\vec x',t')\Psi_a(\vec x,t)  \rangle &  = & -i S^<_{a,b}(\vec x-\vec x';t-t') \label{prop} \\
-i S^<(\vec k ;t-t') & = & \int dq_0 \rho(q_0,\vec k) N(q_0) e^{-iq_0(t-t')} \label{specprop} \\
N(q_0) & = & \frac{1}{e^{\beta (q_0-\mu)}+1} \label{occupa}
\eea

Therefore, we find

\be 
\frac{U(T,\mu)}{V}= {\cal U}(T,\mu) = \int \dbark  \int dq_0 q_0 \mbox{Tr} \left[\gamma^0 \rho(q_0,\vec k) \right]N(q_0) \label{inter}
\ee

In the grand canonical ensemble, the average number of particles is given by 

\be
N = \int d^3 x \langle \bar{\Psi}(\vec x,t) \gamma^0 \Psi(\vec x,t)\rangle = \int d^3 x \mbox{Tr} \left[-i S^<(\vec x-\vec x';t-t') \gamma^0 \right]_{\vec x=\vec x';t=t'} 
\ee
\noindent which is independent of temperature and given in terms of the spectral density by 

\be
N = \int \dbark \int dq_0 \mbox{Tr}\left[ \rho(q_0,\vec k) \gamma^0 \right] N(q_0)
\ee
 
The temperature dependence of the chemical potential $\mu(T)$ is found by requiring that $N$ be independent of temperature.

In the degenerate case temperature only affects states that are a distance $\approx T$ from the Fermi surface, for positive
baryochemical potential only quasiparticle states are important, since the contribution of the antiparticles will
be suppressed by a factor $\exp{-\frac{\mu}{T}} \ll 1$\cite{pisarski,rischke}.  For massless quarks the spectral function $\rho(q_0,\vec k)$ is of the
form\cite{pisarski,rischke}

\be
\rho(q_0,\vec k) = \frac{1}{2} \left[{\cal P}_-(\hat{\bf k})\rho_-(q_0,\vec k) + {\cal P}_+(\hat{\bf k})\rho_+(q_0,\vec k)  \right]
\ee 

\noindent with $ {\cal P}_{\pm} = \gamma^0 \pm \vec{\gamma}\cdot \hat{\bf k}$ and $\rho_{\pm}$ are the spectral density for
quasi-antiparticles (+) and quasiparticles (-) respectively. As argued above temperature affects only states a distance
 $\approx T$ from the Fermi surface, hence only quasiparticle states are important, and antiparticle states  are exponentially suppressed. Therefore the contribution to the internal energy, specific heat, and (quasi) particle density
for states near the Fermi surface are given by

\bea
&&{\cal U}(T,\mu) = 2 \int  dq_0 q_0 \eta(q_0) N(q_0) \label{interener} \\
&& C_v = \frac{d {\cal U}}{dT} \label{Cv} \\
&& {\cal N}= \frac{N}{V} = 2 \int  dq_0  \eta(q_0) N(q_0) \label{partdens}  
\eea
 
\noindent where we have introduced the {\em single quasiparticle density of states}

\be
\eta(q_0) = \int \dbark  \rho_-(q_0,\vec k) \label{densofstates} 
\ee

In perturbation theory, the quark self-energy features an infrared divergence for states near the Fermi energy\cite{son,ren} which lead to the breakdown of the Fermi liquid picture. These divergences are a consequence of the exchange of dynamically screened soft transverse gluons, while longitudinal gluons are Debye screened. In reference\cite{nflqcd} the leading logarithmic infrared divergences were resummed via the renormalization group at $T=0$, leading to a scaling form of the spectral density for frequency and momenta near the Fermi
 surface. The calculation for $T \neq 0$ but
in the degenerate case $T \ll \mu$ is performed in a straightforward manner following the steps in this reference\cite{nflqcd}. At finite temperature and chemical potential the self-energy for transverse gluons receives contributions from both the gluon and fermion  thermal and dense loops\cite{brapis}-\cite{manuelnew}. In the degenerate case derivatives of the Fermi-Dirac distribution function are sharply peaked at the Fermi surface. For $N_c$  colors and $N_f$ quarks in the fundamental representation  and $q_0, k \sim \mu $ and after a renormalization group resummation of the leading logarithmic divergences along the same lines as that in\cite{nflqcd} we find that the improved quasiparticle spectral density for states near the Fermi surface is  given by

\bea
&&\rho_-(q_0,\vec k) = \frac{\sin[\pi \lambda]}{\pi} \frac{|\tilde{q}_0|\left| 
\frac{\tilde{q}_0}{M}\right|^{-2\lambda}}{\left[\left(\tilde{q}_0
\left|\frac{\tilde{q}_0}{M}\right|^{-2\lambda}-\tilde{k}\cos[\pi \lambda]  \right)^2+\left(\tilde{k}\sin[\pi \lambda] \right)^2 \right]   } \label{specdens} \\
&&\tilde{q}_0 = q_0 -\mu ~~; ~~ \tilde{k}= k-\mu \label{difs} \\
&&\lambda = \frac{\alpha_s}{6\pi} \frac{N^2_c-1}{2N_c} \label{lambda} \\
&&M^2(T) = \frac{\alpha_s}{2\pi} N_f \left[\mu^2+ \frac{\pi^2}{3} T^2\right]+ \frac{\pi \alpha_s}{3}\left(\frac{N^2_c-1}{2N_c} \right)T^2 \label{massa}  
\eea

\noindent where the Landau damping scale $M^2(T)$ displays the contribution from quark loops (first term) and gluon loops
(second term)\cite{lebellac}. This form of the spectral density  is valid for $|\tilde{q}_0|;|\tilde{k}| \ll  M$ since the renormalization group resummation only includes the leading infrared logarithmic divergences and neglects perturbative analytic contributions. Since $ T \ll M$ this form of the quasiparticle spectral
density determines the finite temperature properties in the degenerate case.

As discussed in detail in ref.\cite{nflqcd}, this quasiparticle spectral density features a narrow resonance for $\tilde{k}\neq 0$, 
at\cite{nflqcd}

\be
\tilde{q}_0(\tilde{k}) = \mbox{sign}(\tilde{k}) \left[|\tilde{k}| M^{-2\lambda}\cos[\pi \lambda]  \right]^{\frac{1}{1-2\lambda}}
\label{quasidisp}
\ee

Near the position of the resonance the spectral density can be approximated by
a Breit-Wigner form\cite{nflqcd}

\bea
&& \rho_-(q_0,k)|_{ \tilde{q}_0,k\approx \mu} = Z_p[\tilde{k}]\frac{\cos[\pi\lambda]}{\pi} ~ \frac{\Gamma(\tilde{k})}{(\tilde{q}_0-\tilde{q}_0(\tilde{k}))^2+\Gamma^2(\tilde{k})} \label{breitwigner} \\
&& Z_p[\tilde{k}] = \frac{\left|\frac{\tilde{q}_0(\tilde{k})}{M} \right|^{2\lambda}}{(1-2\lambda)} \label{Zp} \\
&& \Gamma(\tilde{k}) = Z_p[\tilde{k}] |\tilde{k}| \sin[\pi \lambda] \label{width}
\eea

Furthermore the group velocity near the Fermi surface is given by

\begin{equation}
v_g(\tilde{k}) =\frac{d\tilde{q}_0}{d\tilde{k}} = Z_p[\tilde{k}]\cos[\pi \lambda] \label{groupvel}
\end{equation}


Therefore the residue of the ``quasiparticle pole'',  the ``quasiparticle width'' and the ``quasiparticle group
velocity'' vanish near the Fermi surface as\cite{nflqcd}

\bea
&&Z_p[\tilde{k}] \propto |k-k_F|^{\frac{2\lambda}{1-2\lambda}} \label{znear} \\
&& \Gamma(\tilde{k})\propto |k-k_F|^{\frac{1}{1-2\lambda}}\label{widthnear} \\
&& v_g(\tilde{k}) = Z_p[\tilde{k}]\cos[\pi \lambda] \propto |k-k_F|^{\frac{2\lambda}{1-2\lambda}} \label{vgroup}
\eea

The vanishing of the residue at the quasiparticle pole  at the Fermi surface is the hallmark of the breakdown of Fermi liquid 
theory.  In a degenerate Fermi gas, all of the thermodynamic response functions are determined by the quasiparticle properties near the Fermi surface. The vanishing of the quasiparticle residue at the Fermi surface, and the consequent breakdown of Fermi liquid theory has the potential for modifying the thermodynamic response functions. The important aspect that we seek to understand is how
these non-Fermi liquid features affect the specific heat of quark matter which is relevant for the cooling of neutron stars with a normal component of quark matter in the core. 

As discussed in\cite{nflqcd} the non-Fermi liquid corrections to the quasiparticle spectral density are {\em non-perturbative} as explicitly displayed by the anomalous powers in the dispersion relation and the wave function renormalization. The renormalization group improved quasiparticle spectral density (\ref{specdens}) has been obtained by a resummation of the leading infrared divergences near the Fermi surface and neglects perturbative corrections which are analytic in the frequency and momentum near the Fermi surface\cite{nflqcd}. The validity of
(\ref{specdens}) is restricted to a region $ -M < \tilde{q}_0;~\tilde{k} < M$ where the contribution from the dressed transverse gluons is dominated by Landau damping\cite{nflqcd}. Away from the Fermi surface the spectral density must match smoothly to that of the
fermionic collective modes (see\cite{nflqcd,blaizot,vanderheyden,manuelbellac}) which for frequency and momenta $>>M$ lead to perturbative corrections to the quasiparticle spectral density for states with frequency and momenta away from the Fermi surface. 

Thus we will restrict our study of the specific heat to the {\em non-perturbative} aspects of the non-Fermi liquid corrections by focusing 
on the contribution to the specific heat of states within a width of order $M$ near the Fermi surface and on the terms with  anomalous exponents which are the hallmark of non-Fermi liquid corrections. Therefore, we will neglect the contributions from perturbative terms  since these will arise from:
i) perturbative corrections to the quasiparticle spectral density not included in (\ref{specdens}) and ii) the contribution from states a distance $>M$ below the Fermi surface, none of which are associated with the breakdown of the Fermi liquid description.

\subsection{Single quasiparticle density of states}

The expressions for the internal energy, specific heat and density given by (\ref{interener}-\ref{partdens}) require the computation of
the single quasiparticle density of states (\ref{densofstates}). The non-Fermi liquid features of the single quasiparticle density of states can be extracted by restricting the momentum integral to states within a width of order $M$ around the Fermi surface, hence non-Fermi liquid contributions to $\eta(q_0)$ are given by    

\be
\eta(q_0) = \frac{1}{2\pi^2} \int^M_{-M} d\tilde{k} \left[\mu + \tilde{k} \right]^2 \rho_-(q_0,\vec k) \label{denso}
\ee

Straightforward integration leads to the following explicit form

\bea
\eta(\tilde{q}_0) &=& \left( {M \over \pi}\right)^2 {\sin \pi \lambda \over \pi}
\left\{ z - z\left({\mu \over M} \sigma + z \; \cos \pi \lambda
\right) \mbox{ArgTh}\left( { 2 \; z \; \cos \pi \lambda \over 1 + z^2
}\right) \right. \cr \cr
&+& \left. { 1 \over 2 \sin \pi \lambda} \left[\left({\mu \over M}
\right)^2 + 2 z {\mu \over M} \sigma \cos \pi \lambda + z^2 \cos 2 \pi \lambda
\right] \left[ \pi - \arctan\left({ 2 \; z \; \sin \pi \lambda \over 1 - z^2
}\right) \right]\right\}\label{densoappx} \eea

\noindent with 

\be
 z \equiv  \left( {|{\tilde q }_0|\over M} \right)^{1 - 2
\lambda} ~~;~~   \sigma \equiv  \mbox{sign}({\tilde q }_0).
\ee

Since the specific heat is determined by a small region of width $ \tilde{q}_0 \approx T \ll M $ we 
approximate the above form of $\eta(\tilde{q}_0)$ for $z \ll 1$

\bea
\eta(\tilde{q}_0) &=& \frac12 \left( {\mu \over \pi}\right)^2 + {M \over \pi^2} 
\; z \left[ \left(1 -{\mu^2 \over M^2} \right){M \over \pi} \sin \pi
\lambda + \mu \; \sigma \cos \pi \lambda \right] \cr \cr
&-& z^2 \; {M \over
\pi^2}  \left[ {2 \mu \sigma \over \pi}   \sin 2 \pi \lambda - {M
\over 2} \cos 2 \pi \lambda \right] + {\cal O}(z^3) \label{densosmallz}
\eea

This expression has the correct free-field limit, which corresponds to $\rho_-(q_0,\vec k) = \delta(q_0-k)$.
A noteworthy point to emphasize is that while the quasiparticle residue vanishes at the Fermi surface, the single quasiparticle
density of states is non-vanishing and {\em finite} but with a divergent frequency derivative   at the Fermi surface ($\tilde{q}_0=0$). 
 This divergence in the frequency derivative  introduces non analytic  finite
temperature corrections to the chemical potential which are explored below. 

\subsection{Chemical Potential:} 

In the grand canonical ensemble, the temperature dependence of the chemical potential is obtained by requiring that the mean number of particles is  {\em fixed} and is independent of temperature. The chemical potential is the Lagrange multiplier that enforces the constraint of a given particle number and  its  temperature dependence is obtained from the condition

\be
\frac{d{\cal N}}{dT}=0 = \int dq_0 \left[\eta(q_0)\frac{dN(q_0)}{dT}+\frac{d \eta(q_0)}{dT} N(q_0) \right] \label{chemicond}
\ee  

\noindent with ${\cal N}$ is the particle density given by (\ref{partdens}). 

 In the degenerate case ${dN(q_0)}/{dT}$ is localized in a region of width $\sim T$ around $q_0 =\mu$ hence the
first term in (\ref{chemicond}) above will only receive contributions from quasiparticles near the Fermi surface for which the single
quasiparticle density of states is given by (\ref{densoappx}). 

While in the free theory the single particle density of states does not depend on the temperature or chemical potential, in the 
interacting theory it depends on $T$ both through $M(T)$ given by (\ref{massa}) as well as through the chemical potential, thus the second term in (\ref{chemicond}) is non-vanishing. However, while the
second term in (\ref{chemicond}) receives contributions from states well below the Fermi surface, the non-perturbative non-Fermi liquid corrections are only associated with states within a region of width $M$ near the Fermi surface. Quasiparticle states or collective modes well below the Fermi surface give  perturbative corrections which are analytic in $\tilde{q}_0$ in the region $\tilde{q}_0 /M \ll 1$. Such contributions will be neglected in this article.

The contribution to $d\eta/dT$ from 
the quasiparticle states near the Fermi surface given by eq.(\ref{densoappx}) can be obtained  as follows.
Writing

\be
\frac{d\eta(\tilde{q}_0)}{dT} = \frac{\partial \eta(\tilde{q}_0)}{\partial \mu}\frac{d \mu}{dT}+ \frac{\partial \eta(\tilde{q}_0)}{\partial M}\frac{d M}{dT} \label{detadT}
\ee
 
\noindent it is readily apparent from eq. (\ref{densoappx}) or its small $z$ approximation eq.(\ref{densosmallz}) that

\be
\frac{\partial \eta(\tilde{q}_0)}{\partial M} \propto \lambda ~~; ~~ \frac{d M}{dT} \propto \lambda \frac{T}{M}
\ee

\noindent therefore the second contribution in eq. (\ref{detadT}) is perturbatively small ${\cal O}(\lambda^2 T/M)$, does not lead to anomalous powers of the ratio $T/M \ll 1 $  and will be neglected, consistently with neglecting perturbative
corrections in the spectral density and the quasiparticle density of states. Upon integrating in $\tilde{q}_0$ the term
$\frac{\partial \eta(\tilde{q}_0)}{\partial \mu}N(\tilde{q}_0)$  up to a cutoff of order $\sim M$,
one finds that the contribution from the first term of (\ref{detadT}) is again perturbative, and of the same order as
terms that have been neglected in the spectral and quasiparticle densities. Therefore, we conclude that the contribution of the second term
in eq. (\ref{chemicond}) is of the same order as terms that were neglected in the perturbative expansion and hence it will be consistently neglected.

 Hence, we are led to the following  temperature derivative of the chemical potential to leading order

\be
\frac{d\mu(T)}{dT} = - \frac{\int^M_{-M} d\tilde{q}_0 \eta(\tilde{q}_0) D(\tilde{q}_0)(\beta \tilde{q}_0)}{\int^M_{-M} d\tilde{q}_0 \eta(\tilde{q}_0) D(\tilde{q}_0)} ~~; ~~ D(x)= \frac{e^x}{\left[e^x+1\right]^2}\label{chemicalpotder}
\ee

The leading contribution in the weak coupling and degenerate limit can be extracted by changing variables $\beta \tilde{q}_0=x$ and taking $M/T \rightarrow \infty$. We find 

\be
{ d \over dT} \mu(T) = - 4 \, \cos \pi \lambda\; F(\lambda) \left(\frac{T}{\mu(T)}\right) \left(\frac{T}{M}\right)^{ - 2 \,
\lambda} + {\cal O}\left(\left(\frac{T}{\mu(T)}\right)^2\left(\frac{T}{M}\right)^{ - 4 \,
\lambda}\right)\label{deriT}
\ee

\noindent leading to the following form of the temperature dependence of the chemical potential

\be
\mu^2(T) = \mu^2(0)\left[1 - \kappa(\lambda) \; \left(\frac{T}{\mu(0)}\right)^2 \left(\frac{T}{M(0)}\right)^{ - 2 \,
\lambda} + {\cal O}\left(\left(\frac{T}{\mu(0)}\right)^3 \left(\frac{T}{M(0)}\right)^{ - 4 \,
\lambda}\right)\right]
\ee

\noindent where we have introduced the functions

\bea
F(\lambda) &=&   \; \Gamma(3 - 2 \, \lambda)
\; \zeta(2 - 2 \, \lambda) \; (1 - 2^{ 2 \, \lambda-1}) \label{F} \\
\kappa(\lambda) &=& 8 \, \cos( \pi \lambda) \; \Gamma(2 - 2 \, \lambda)
\; \zeta(2 - 2 \, \lambda) \; (1 - 2^{ 2 \, \lambda-1}) = { 4 \, \cos
\pi \lambda \over 1-\lambda } \; F(\lambda) \label{K}
\eea

\noindent in terms of the Gamma  and Riemann zeta functions.

The non-analiticity of the temperature derivative of the chemical potential reflects the divergence of the frequency
derivative of the single quasiparticle density of states near the Fermi surface.

\subsection{Specific Heat:} 

The specific heat (per unit volume) given by (\ref{Cv}) can be written as

\be
C_v= 2 \int dq_0 q_0  \left[\eta(q_0)\frac{dN(q_0)}{dT}+\frac{d \eta(q_0)}{dT} N(q_0) \right] \label{CV1}
\ee

To extract the leading order in non-Fermi liquid corrections we follow the analysis invoked in obtaining the
temperature derivative of the chemical potential. In the degenerate case the first term in eq. (\ref{CV1}) above is determined by the quasiparticle density of states eq. (\ref{densoappx}) for $\tilde{q}_0 \sim T \ll M$. An estimate
of the second term is obtained by following the arguments presented for the case of the chemical potential. A similar
analysis reveals that the second term is perturbative in the weak coupling limit and of the same order as terms that
were neglected in obtaining the spectral density and the single quasiparticle density of states near the Fermi surface.
Therefore the second term in eq. (\ref{CV1}) will be consistently neglected. 

Using the condition (\ref{chemicond})
and the fact that near the Fermi surface the quasiparticle density of states is a function of $\tilde{q}_0$, the leading order (in coupling)
contribution to the specific heat is obtained from 

 \be
C_v= 2 \int^M_{-M} d\tilde{q}_0 \tilde{q}_0  \eta(\tilde{q}_0)\beta D(\tilde{q}_0)\left[\beta \tilde{q}_0 + \frac{d\mu(T)}{dT} \right] \label{Cvsimple}
\ee

\noindent with $D(x)$ given in eqn. (\ref{chemicalpotder}). 

After a rescaling $\tilde{q}_0=xT$,  using the result (\ref{deriT}) for $\frac{d\mu(T)}{dT}$ and neglecting terms perturbative in the coupling $\lambda$  we  find

\bea
C_v & = &    \frac{\mu^2(0) T}{3} \left[1+   { 6 \over \pi^2}  \left[ F(2 \, \lambda-1) \, \cos 2 \pi \lambda +
8 \,  F(\lambda)^2 \; \cos^2 \pi \lambda \right]  \left(\frac{T}{\mu(0)}\right)^2 \left(\frac{T}{M(0)}\right)^{ - 4 \, \lambda} \right. \nonumber \\
&-& \left.  \kappa(\lambda)\;\left(\frac{T}{\mu(0)}\right)^2 \left(\frac{T}{M(0)}\right)^{ - 2 \,
\lambda} + {\cal O}\left( \left(\frac{T}{\mu(0)}\right)^3\left(\frac{T}{M(0)}\right)^{ - 6 \,
\lambda}\right)\right]
\eea

\noindent with the functions $F(\lambda)~;~\kappa(\lambda)$ given by eq. (\ref{F},\ref{K}) above.

 The non-Fermi liquid contributions are explicit in the anomalous power laws with
the temperature, which are a result of the divergent frequency derivative of the single quasiparticle density of
states near the Fermi surface. We emphasize that for $T \ll M$ the non-fermi liquid corrections to the specific heat are
much larger than those of the massless gluons or any other Goldstone bosons to the specific heat, which is of order $T^3$. 

To illustrate the magnitude of the non-Fermi liquid corrections to the specific heat we show in fig. 1 the function 
$G(z)= 3 C_v / \mu^2(0)T$ vs. $z=T/\mu(0)$ for three colors and two flavors of massless quarks. We have used the one loop QCD running coupling  at a scale $\mu(0)$ without further justification\footnote{Reference\cite{bedaque} gives a thorough discussion of the dependence of the coupling on the chemical potential and the inherent ambiguities in the choice of scale.}. The solid line corresponds to $\mu(0)=0.5~\mbox{Gev}$ and the dashed line to $\mu(0)=1~\mbox{Gev}$, for these 
values of the strong coupling constant, the
value of the effective coupling at these scales are $\lambda= 0.152$ for $\mu =0.5~\mbox{Gev}$ and $\lambda = 0.046$ for
$\mu = 1~\mbox{Gev}$ therefore the reliability of the perturbative calculation for the values of the chemical potentials expected at the core of neutron stars is at best questionable. This issue notwithstanding this qualitative estimate points out that non-Fermi liquid corrections to the specific heat result in a change of a few percent. The physical reason for this result is that whereas the quasiparticle residue (wave function renormalization) 
{\em vanishes} at the Fermi surface, the single quasiparticle density of states is non-vanishing and finite, but its 
frequency derivative at the Fermi surface is divergent. The finite, non-zero value of the density of states implies
that the non-Fermi liquid features will only result in corrections to the Fermi liquid specific heat. The numerical estimate, with the caveat that its regime of validity is restricted to small coupling, indicates that these corrections are
rather small.  

Keeping higher order terms in the perturbative expansion, will not dramatically modify this conclusion since our analysis captures the most relevant aspects of the non-Fermi liquid corrections. For the values of temperature and chemical potential corresponding to young neutron stars we see that the non-fermi liquid corrections amount to a few percent, unlikely to make a dramatic change in the cooling history of young neutron stars. For protoneutron stars
with higher temperatures $T \sim 30-50 ~\mbox{Mev}$ and smaller chemical potentials, the corrections could be of order
$10-15 \%$ but in this regime the reliability of the calculation is questionable. We emphasize that this numerical analysis is only intended to provide a qualitative estimate of the magnitude of the non-Fermi liquid corrections but should not be taken as a quantitative indicator since higher order terms, which are subleading and have been neglected in the perturbative expansion can become comparable in this range of couplings. 

\subsection{Discussion}

The fact that the wave function renormalization, i.e, the residue at the ``quasiparticle pole'' vanishes at the
Fermi surface as given by eqn. (\ref{znear}) would seem to indicate that quasiparticle density of states should
vanish near the Fermi surface with an anomalous power law. This in turn would suggest that the linear power of
temperature, which is the leading contribution in the degenerate limit, would be multiplied by an 
anomalous power of $T/\mu$. However, the reason
that the quasiparticle density of states remains {\em finite} at the Fermi surface despite the vanishing of the
quasiparticle pole is because the group velocity vanishes proportionally to the wave function renormalization as
displayed by eqn. (\ref{vgroup}).  To see this feature clearly, we can approximate the spectral density (\ref{breitwigner})
in the limit of vanishing width by 

\be
\rho_-(q_0,k)_{\tilde{k}\approx 0} \sim Z_p[\tilde{k}] \cos[\pi \lambda] \delta(\tilde{q}_0- \tilde{q}_0(\tilde{k}))\label{narrowidth}
\ee
\noindent with $\tilde{q}_0(\tilde{k})$ given by eqn. (\ref{quasidisp}). The integral over $\tilde{k}$ in eqn. (\ref{denso})
can be performed by changing variables to $\tilde{q}_0(\tilde{k})$ where the Jacobian of the transformation is precisely
$1/v_g$ with $v_g$ the group velocity and by  eqn. (\ref{vgroup}) this Jacobian {\em exactly cancels} the wave-function
renormalization. This analysis clearly reveals that the reason that the quasiparticle density of states at the Fermi
surface remains finite is  the cancellation between the residue at the quasiparticle pole and the group
velocity, which both vanish proportionally near the Fermi surface. 

At this stage it is also important to compare our results to those obtained in reference\cite{norton} in which
the infrared divergences associated with magnetic photon exchange were studied for the first time within the
context of the non-relativistic electron gas. There are several important differences between our treatment and results
and those of reference\cite{norton}. i) In reference\cite{norton} the temperature derivative of the chemical potential
was {\em neglected}  in the derivative of the grand potential with respect to temperature (see equations (5.6) and (5.7)
of reference\cite{norton}). This is an important source of
discrepancy because the temperature derivative of the chemical potential has a linear term in the temperature that
cancels against a similar one arising from derivatives of the density of states and the distribution function. We have
included the temperature derivative of the chemical potential in the computation of the specific heat explicitly
through the condition (\ref{chemicond}). ii) The calculation in \cite{norton} missed the fact that the vanishing
of the group velocity is compensated by the vanishing of the residue at the quasiparticle pole as explicitly discussed above. 
iii) While in reference~\cite{norton} only the perturbative logarithmic terms were accounted for, the renormalization
group resummation makes clear that these logarithms sum up to anomalous power laws, a naive expansion in terms of
the coupling constant reveals those logarithms, but the resummation makes explicit the {\em vanishing of both the 
residue and the group velocity}. 

A consistent resummation and analysis of the quasiparticle pole and group velocity as explicitly discussed here reveals that
the density of states remains finite at the Fermi surface and the non-Fermi liquid corrections are subleading.  


\section{Conclusions}

The exchange of dynamically screened gluons leads to a pairing instability and at the same time to a breakdown of the
Fermi liquid description of the normal state of QCD. The superconducting component does not contribute to the cooling
of young neutron stars because of the exponential suppression of the specific heat and the neutrino emissivity. If 
color superconductivity is in the form of a 2SC condensate, or some quarks do not pair, the normal quarks contribute to
cooling of the neutron star via neutrino and antineutrino emission.  

The search for novel observational aspects of quark matter at the core
of neutron stars motivated us to study the effects of the breakdown of the Fermi liquid description of quasiparticle
states of the normal component near the Fermi surface in the specific heat. This aspect is relevant to understand the  cooling rate through  direct quark Urca processes. Our analysis is based on  a renormalization group improvement of the single quasiparticle spectral
density in the degenerate case and reveals that non-Fermi liquid corrections tend to decrease  the specific heat and therefore increase the cooling rate. However these corrections  are rather small, about $10-15\%$ even in the most favorable scenarios,  and probably
unobservable in the cooling history of neutron stars or even protoneutron stars. A complete assessment of potential non-Fermi liquid corrections to the cooling
rate still requires an understanding of these corrections on the neutrino emissivity. Both quarks and leptons (electrons)
that enter in the quark direct Urca process will receive non-Fermi liquid corrections through their spectral density
although the corrections for quarks, in terms of the strong coupling constant are more important than those for the electrons. We are currently studying these corrections in the weak matrix elements for the quark direct Urca processes and
expect to report on our results in the near future.

\acknowledgements
D.B. thanks NSF for support through grants PHY-9605186, PHY-9988720 and  NSF-INT-9815064.
H. J. d. V. thanks the CNRS-NSF collaboration for support. The authors  thank K. Rajagopal and M. Alford for 
correspondence and comments.




\begin{figure}
\centerline{ \epsfig{file=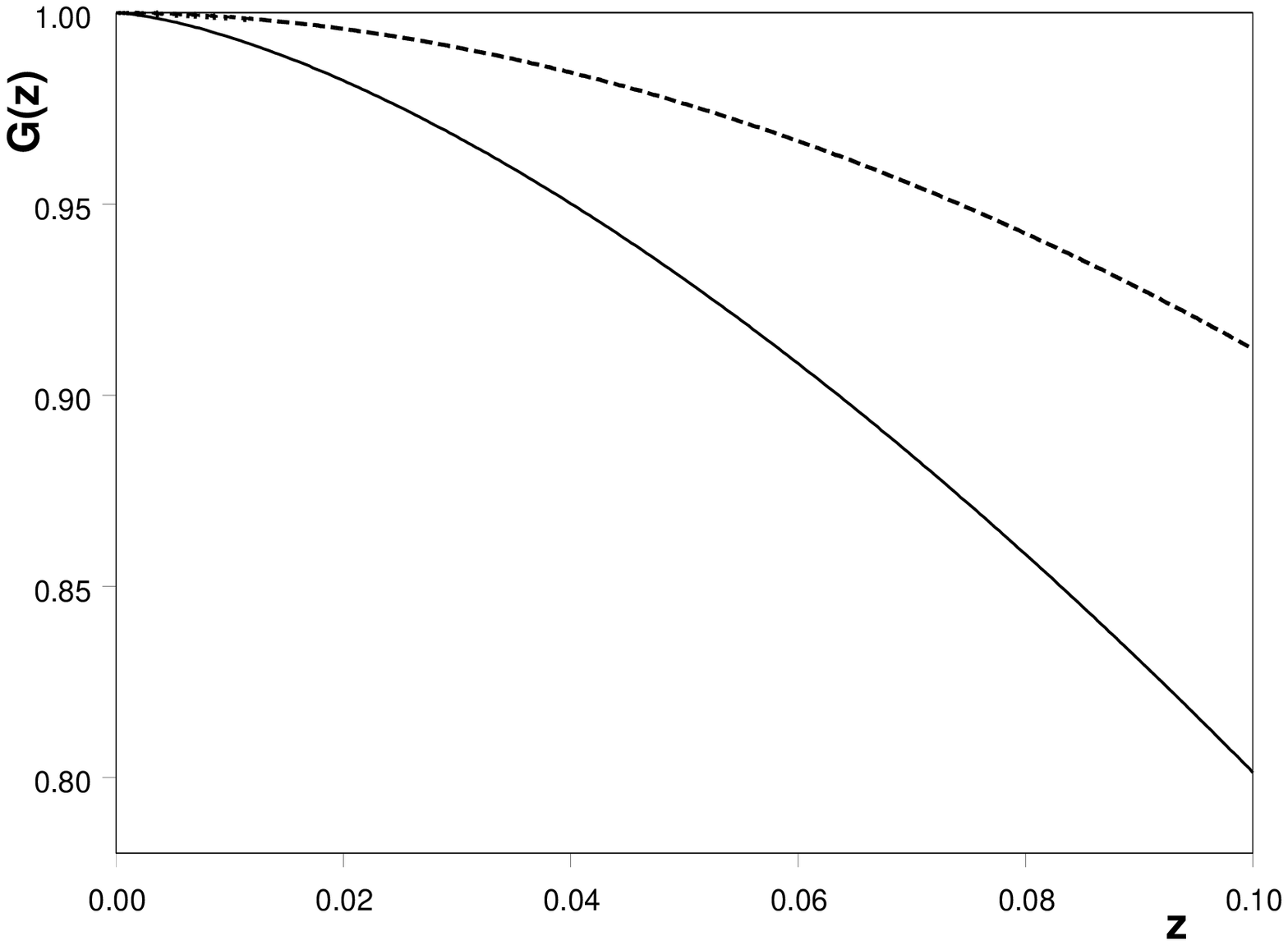,width=5in,height=5in}}
\caption{$G(z)=\frac{3C_v}{\mu^2(0)T}$ vs. $z=\frac{T}{\mu(0)}$ for $N_c=3;N_f=2$. Solid line: $\mu(0)=0.5~\mbox{Gev}$, dashed line: $\mu(0)=1~\mbox{Gev}$.\label{fig3}} 
\end{figure}



\end{document}